\documentclass[showpacs,preprintnumbers,12pt,onecolumn]{revtex4}
\usepackage{amssymb}
\usepackage{amsfonts}
\usepackage{amsmath}
\usepackage{graphicx}
\usepackage{dcolumn}
\usepackage{bm}

\input{tcilatex}

\begin{document}

\title{Universality in ratchets without spatial asymmetry}
\author{Ricardo Chac\'{o}n}
\affiliation{Departamento de F\'{\i}sica Aplicada, Escuela de Ingenier\'{\i}as
Industriales, Universidad de Extremadura, Apartado Postal 382, E-06071
Badajoz, Spain}
\date{\today}

\begin{abstract}
It is demonstrated that to optimally enhance directed transport by symmetry
breaking of temporal forces there exists a universal force waveform which
allows to deduce universal scaling laws that explain previous results for a
great diversity of systems\ subjected to a standard biharmonic force and
provide a universal quantitative criterion to optimize any application of
the ratchet effect induced by symmetry breaking of temporal forces.
\end{abstract}

\maketitle





\textbf{\ }Understanding the ratchet effect\ [1-4] induced by symmetry
breaking of temporal forces is a fundamental issue that has remained
unresolved for decades. While the dependence of the directed transport on
each of the ratchet-controlling parameters has been individually
investigated, there is still no general criterion to apply to the whole set
of these parameters to optimally control directed transport in general
systems without a ratchet potential [5-25]. Consider a general deterministic
system (classical or quantum, dissipative or non-dissipative, one- or
multi-dimensional) subjected to a $T$-periodic zero-mean ac force $f\left(
t\right) $ where a ratchet effect is induced by solely violating temporal
symmetries. A popular choice would be the simple case of a biharmonic force, 
$f_{h1,h2}(t)=\epsilon _{1}har_{1}\left( \omega t+\varphi _{1}\right)
+\epsilon _{2}har_{2}\left( 2\omega t+\varphi _{2}\right) $, where $har_{1,2}
$ represents indistinctly $\sin $ or $\cos $. Clearly, the aforementioned
symmetries are solely the shift symmetry of the force $\left( f\left(
t\right) =-f\left( t+T/2\right) ,T\equiv 2\pi /\omega \right) $ and the
time-reversal symmetry of the system's dynamic equations. Of course, the
breaking of the latter symmetry implies the breaking of some time-reversal
symmetry of the force $\left( f\left( -t\right) =\pm f\left( t\right)
\right) $ in some general case, but not in all cases [19]. The analysis of
the breaking of these two fundamental symmetries allows to find the regions
of the parameter space $\left( \epsilon _{1},\epsilon _{2},\varphi
_{1},\varphi _{2}\right) ,\epsilon _{1}+\epsilon _{2}=const$., where the
ratchet effect is optimal in the sense that the average of relevant
observables (such as velocity and current, hereafter referred to as $%
\left\langle V\right\rangle $) is maximal, the remaining parameters being
held constant. In this Letter, it is shown that such regions are those where
the \textit{effective} degree of symmetry breaking is \textit{maximal}. The
theory arises from the observation that Curie's principle [26] implies that
a broken symmetry is a structurally stable situation [4]. At this point a
quantitative measure of the degree of symmetry breaking (DSB) is introduced,
on which the strength of directed transport must depend. This quantitative
relationship between cause (symmetry breaking) and effect (directed
transport) is hereafter referred to as the DSB mechanism. Also, this
quantitative relationship is expected to exhibit a dependence on the
symmetry-breaking parameters which is universal if and only if the symmetry
breaking takes place solely in the driving force, i.e., in the external
agent which is simultaneously the transport-inducing force and the
ratchet-inducing force. Since the ratchet effect can occur at any
spatio-temporal scale, such a quantitative measure of the DSB must be
independent of the force's amplitude and period. I define consistently the
DSB of the symmetries of the force $f(t)$ by the expressions%
\begin{eqnarray}
D_{s}(f) &\equiv &\left\langle \frac{-f\left( t+T/2\right) }{f(t)}%
\right\rangle _{T}\equiv \frac{1}{T}\int_{0}^{T}\frac{-f\left( t+T/2\right) 
}{f(t)}dt,  \notag \\
D_{\pm }(f) &\equiv &\left\langle \frac{\pm f\left( -t\right) }{f(t)}%
\right\rangle _{T}\equiv \frac{1}{T}\int_{0}^{T}\frac{\pm f\left( -t\right) 
}{f(t)}dt,  \TCItag{1}
\end{eqnarray}%
where increasing deviation of $D_{s,\pm }(f)$ from 1 (unbroken symmetry)
indicates an increase in the DSB. But the effectiveness of any periodic
zero-mean force at producing transport diminishes as the transmitted impulse
over a half-period is decreased while its amplitude and period are held
constant. In general, this means that optimal enhancement of the ratchet
effect is achieved when maximal \textit{effective} symmetry breaking occurs,
which is in turn a consequence of two reshaping-induced competing effects:
the increase of the degree of breaking of the force's symmetries and the
decrease of the transmitted impulse over a half-period, thus implying the
existence of a universal force waveform which optimally enhances the ratchet
effect. Thus, for the biharmonic force $f_{\cos ,\cos }\left( t\right)
=\epsilon _{1}\cos \left( \omega t+\varphi _{1}\right) +\epsilon _{2}\cos
\left( 2\omega t+\varphi _{2}\right) $, Eq. (1) can be put into the form $%
D_{s}(f_{\cos ,\cos })=1-\left( a/\pi \right) \int_{0}^{2\pi }\cos \left(
2\tau +\varphi _{eff}\right) /P\left( \tau ;a,\varphi _{eff}\right) d\tau $, 
$D_{+}(f_{\cos ,\cos })=1+\left( a/\pi \right) \int_{0}^{2\pi }\sin \left(
2\tau \right) \sin \varphi _{eff}/P\left( \tau ;a,\varphi _{eff}\right)
d\tau $, $D_{-}(f_{\cos ,\cos })=1-\left( 1/\pi \right) \int_{0}^{2\pi }%
\left[ \cos \tau +a\cos \left( 2\tau \right) \cos \varphi _{eff}\right]
/P\left( \tau ;a,\varphi _{eff}\right) d\tau $, where $a\equiv \epsilon
_{2}/\epsilon _{1},\tau \equiv \omega t+\varphi _{1},\varphi \equiv \varphi
_{2}-2\varphi _{1},P\left( \tau ;a,\varphi _{eff}\right) \equiv \cos \tau
+a\cos \left( 2\tau +\varphi _{eff}\right) $. The quantity $\varphi _{eff}$
is hereafter referred to as the \textit{effective phase}$.$These integrals
diverge at the zeros of the quartic polynomial $4a^{2}x^{4}+4a\cos \varphi
x^{3}+\left( 1-4a^{2}\right) x^{2}-2a\cos \varphi x+a^{2}\cos ^{2}\varphi =0$%
, where $x\equiv \cos \tau $. After solving this algebraic equation for $x$,
one obtains that the three integrals diverge when $a\geqslant 1/2,\varphi
_{eff}=\left\{ \pi /2,3\pi /2\right\} $, and thus the DSB is maximal at
these parameter values for the three symmetries of the force (see Fig. 1d).
For these values of the effective phase, one finds that the transmitted
impulse over a half-period is maximal at $a=1/2$ while the biharmonic
force's amplitude is held constant. This means that maximal \textit{effective%
} symmetry breaking occurs at $a=1/2,\varphi _{eff}=\left\{ \pi /2,3\pi
/2\right\} $. A similar analysis of the remaining three versions of the
biharmonic force yields the results summarized in Table I (second column),
which are again the same for the three symmetries in each case. Note that
one could equivalently define a measure of the DSB by taking the time
average of the inverse quantities $-f(t)/f(t+T/2),\pm f(t)/f(-t)$: one finds
that the corresponding measure (1) exhibits the same qualitative behaviour
as a function of the symmetry-breaking parameters, and exactly the same
optimal values of these parameters are found to yield a maximal effective
DSB. This indicates that (1) provides a \textit{bona fide }measure of the
DSB. Remarkably, such optimal parameter values correspond to a \textit{single%
} optimal waveform for the four versions of the biharmonic force (see Fig.
1b). The DSB mechanism implies that such a waveform is universal, i.e., it
corresponds to a force waveform which optimally enhances the ratchet effect
in any system. Consider now the case of the elliptic force $%
f_{ellip}(t)=\epsilon f(t;T,m,\theta )\equiv \epsilon \limfunc{sn}\left(
\Omega t+\Theta ;m\right) \limfunc{cn}\left( \Omega t+\Theta ;m\right) $,
where $\limfunc{cn}\left( \cdot ;m\right) $ and $\limfunc{sn}\left( \cdot
;m\right) $ are Jacobian elliptic functions [27] of parameter $m$, $\Omega
\equiv 2K(m)/T,$ $\Theta \equiv K(m)\theta /\pi $, $K(m)$ is the complete
elliptic integral of the first kind [27], $T$ is the period of the force,
and $\theta $ is the (normalized) initial phase $\left( \theta \in \left[
0,2\pi \right] \right) $. Fixing $\epsilon ,T$, and $\theta $, the force
waveform changes as the shape\textit{\ }parameter $m$ varies from 0 to 1
(see Fig. 1a). In this case, Eq. (1) yields $D_{s}(f_{ellip})=E(m)K^{-1}(m)%
\left( 1-m\right) ^{-1/2}$, where $E(m)$ is the complete elliptic integral
of the second kind [27] (see Fig. 1c). Physically, the motivation for this
choice is that $f_{ellip}(t;T,m=0,\theta )=\epsilon \sin \left( 2\pi
t/T+\theta \right) /2$, and that $f_{ellip}(t;T,m=1,\theta )$ vanishes,
i.e., in these two limits directed transport is not possible, while it is
expected for $0<m<1$. Thus, one may expect in general the average of any
relevant observable $\left\langle V\right\rangle $ to exhibit an extremum at
a certain critical value $m=m_{c}$ as the shape parameter $m$ is varied, the
remaining parameters being held constant. The DSB mechanism implies that
such a value $m_{c}$ is universal, i.e., it corresponds to a universal force
waveform which optimally enhances the ratchet effect in any system. \textit{%
Universality} requires that such an optimal waveform should be closely
related to that deduced for the case of a biharmonic force, in the sense of
its Fourier series. Indeed, using $f_{ellip}(t;T,m,\theta )/\epsilon
=\sum_{n=1}^{\infty }a_{n}(m)\sin \left[ n\left( 2\pi t/T+\theta \right) %
\right] $, $a_{n}(m)\equiv n\pi ^{2}m^{-1}K^{-2}(m)\func{sech}\left[ n\pi
K(1-m)/K(m)\right] $ one could expect the critical value $m_{c}$ to be near $%
m=0.983417$ since $f_{ellip}(t;T,m=0.983417,\theta )/\epsilon =a_{1}\left(
m=0.983417\right) \left[ \sin \left( 2\pi t/T+\theta \right) +\left(
1/2\right) \sin \left( 4\pi t/T+2\theta \right) +0.178592\sin \left( 6\pi
t/T+3\theta \right) +...\right] $, i.e., the optimal values $\left( \epsilon
_{2}/\epsilon _{1}=1/2,\varphi _{2}-2\varphi _{1}=0\right) $ for the
biharmonic approximation of the elliptic function are recovered at $%
m=0.983417$ (cf. Table I, second column, and compare Figs. 1a and 1b).
Numerical studies of diverse systems [28] confirmed the universality and
accuracy of the critical value $m_{c}=0.983...$, i.e., the universality of
the optimal waveform. Similarly, from the Fourier series of a sawtooth-wave
force $f_{sawtooth}\left( t,T\right) /\epsilon =2\left[ \sin \left( 2\pi
t/T\right) -\left( 1/2\right) \sin \left( 4\pi t/T\right) +(1/3)\sin \left(
6\pi t/T\right) -...\right] $, one recovers the optimal values $\left(
\epsilon _{2}/\epsilon _{1}=1/2,\varphi _{2}-2\varphi _{1}=\pi \right) $ for
its biharmonic approximation (cf. Table I, second column), which explains
the great effectiveness of this waveform in controlling directed transport
of magnetic flux quanta [23].

Next, one exploits the aforementioned universality expected from the DSB
mechanism to deduce the dependence of $<V>$ on the symmetry-breaking
parameters $\left( \epsilon _{1},\epsilon _{2},\varphi _{1},\varphi
_{2}\right) $ of the biharmonic force $f_{h1,h2}(t)$ in leading order for
the usual case [5-22,24,25] of small amplitudes $\left( 1/\epsilon
_{1,2}\rightarrow \infty \right) $. For the sake of clarity, consider first
the case where the violation of the time-reversal symmetry of the system's
dynamic equations can be absorbed in the temporal force because dissipation
is negligible and the Lagrangian (Hamiltonian) of the system does not
contain any additional term explicitly breaking the time-reversal symmetry.
This means that the breaking of the time-reversal symmetry implies the
breaking of the force's symmetry $f\left( -t\right) =f\left( t\right) $.
According to the above arguments, one generally expects $\left\langle
V\right\rangle \sim s\left( \epsilon _{1},\epsilon _{2},\varphi _{1},\varphi
_{2}\right) $, where it is assumed without loss of generality that the
function $s$ is $k$-times piecewise continuously differentiable. From
MacLaurin's series, one has $s\left( \epsilon _{1},\epsilon _{2},\varphi
_{1},\varphi _{2}\right) =\sum_{k=0}^{\infty }\sum_{n=0}^{\infty
}c_{k,n}\left( \varphi _{1},\varphi _{2}\right) \epsilon _{1}^{k}\epsilon
_{2}^{n}$ with $c_{k,0}=c_{0,n}=0$ since the shift symmetry is never broken
in the case of a single harmonic function. The transformation $\epsilon
_{i}\rightarrow -\epsilon _{i},i=1,2,$ implies $f_{h1,h2}(t)\rightarrow
-f_{h1,h2}(t)$, and hence $<V>\rightarrow -<V>$. This means that $s\left(
\epsilon _{1},\epsilon _{2},\varphi _{1},\varphi _{2}\right) \rightarrow
-s\left( -\epsilon _{1},-\epsilon _{2},\varphi _{1},\varphi _{2}\right) $
and hence $k+n=2m+1,m=1,2,...$ . Thus, one obtains 
\begin{equation}
s\left( \epsilon _{1},\epsilon _{2},\varphi _{1},\varphi _{2}\right)
=c_{1,2}\left( \varphi _{1},\varphi _{2}\right) \epsilon _{1}\epsilon
_{2}^{2}+c_{2,1}\left( \varphi _{1},\varphi _{2}\right) \epsilon
_{1}^{2}\epsilon _{2}+O\left( \epsilon _{1}^{2}\epsilon _{2}^{3},\epsilon
_{1}^{3}\epsilon _{2}^{2},\epsilon _{1}^{1}\epsilon _{2}^{4},\epsilon
_{1}^{4}\epsilon _{2}^{1}\right) ,  \tag{2}
\end{equation}%
for $\epsilon _{1,2}$ sufficiently small. Since the ratchet effect does not
depend on the time origin, $<V>$ must remain invariant under the
transformation $t\rightarrow t+t_{0},\forall t_{0}$. This transformation
yields $f_{h1,h2}(t)\rightarrow $ $\epsilon _{1}har_{1}\left( \omega t+%
\widetilde{\varphi }_{1}\right) +\epsilon _{2}har_{2}\left( 2\omega t+%
\widetilde{\varphi }_{2}\right) $, with the fundamental property $\widetilde{%
\varphi }_{2}-2\widetilde{\varphi }_{1}=\varphi _{2}-2\varphi _{1}$, i.e.,
the effective phase $\varphi _{2}-2\varphi _{1}$ remains invariant under
time translation (see Table I, third column), and hence%
\begin{equation}
\left\langle V\right\rangle \sim c_{1,2}\left( \varphi _{2}-2\varphi
_{1}\right) \epsilon _{1}\epsilon _{2}^{2}+c_{2,1}\left( \varphi
_{2}-2\varphi _{1}\right) \epsilon _{1}^{2}\epsilon _{2}.  \tag{3}
\end{equation}%
The transformation $\varphi _{i}\rightarrow \varphi _{i}-\pi ,i=1,2,$
implies $f_{h1,h2}(t)\rightarrow -f_{h1,h2}(t)$, and hence $<V>\rightarrow
-<V>$ whereby $c_{1,2}\left( \varphi _{2}-2\varphi _{1}\right)
=-c_{1,2}\left( \varphi _{2}-2\varphi _{1}+\pi \right) ,c_{2,1}\left(
\varphi _{2}-2\varphi _{1}\right) =-c_{2,1}\left( \varphi _{2}-2\varphi
_{1}+\pi \right) $, while the transformation $\epsilon _{1}\rightarrow
-\epsilon _{1},\varphi _{1}\rightarrow \varphi _{1}-\pi $ maintains $%
\left\langle V\right\rangle $ invariant, and hence $c_{1,2}\left( \varphi
_{2}-2\varphi _{1}\right) =-c_{1,2}\left( \varphi _{2}-2\varphi _{1}+2\pi
\right) ,c_{2,1}\left( \varphi _{2}-2\varphi _{1}\right) =c_{2,1}\left(
\varphi _{2}-2\varphi _{1}+2\pi \right) $. The comparison of these four
relationships for the functions of the effective phase implies that $%
c_{2,1}\left( \varphi _{2}-2\varphi _{1}\right) $ is a $2\pi $-periodic
function while $c_{1,2}\left( \varphi _{2}-2\varphi _{1}\right) \equiv 0$.
Thus, Eq. (3) reduces to%
\begin{equation}
\left\langle V\right\rangle \sim \left( 1/\epsilon _{1}\right) ^{-2}\left(
1/\epsilon _{2}\right) ^{-1}c_{2,1}\left( \varphi _{eff}\right) ,  \tag{4}
\end{equation}%
where a \textit{power law} for the dependence on the amplitudes is now
explicit. In this regard, it is worth noting the great similarity between
the present theory and the highly optimized tolerance (HOT) theory [29]
where a power law is generated by the actions of an external agent aiming to
optimize the behaviour of a system. However, we have seen above that
universality comes from \textit{criticality} in the present theory, while
for HOT systems the details matter. To obtain an explicit expression for the
function $c\left( \varphi _{eff}\right) $ it is useful to consider the
general transformation $t\rightarrow -t+t_{0},\forall t_{0}$. This
transformation yields $f_{h1,h2}(t)\rightarrow $ $\epsilon _{1}har_{1}\left(
\omega t+\widetilde{\varphi }_{1}\right) +\epsilon _{2}har_{2}\left( 2\omega
t+\widetilde{\varphi }_{2}\right) $ where the effective phase is no longer
strictly invariant but changes according to Table I (fourth column). Since
the transformation $t\rightarrow -t$ implies $<V>\rightarrow -<V>$ when the
time-reversal symmetry is unbroken, the change rules of the effective phase
imply that the function $c_{2,1}\left( \varphi _{eff}\right) $ has
necessarily a definite parity (cf. Eq. (4) and Table I, fourth column).
Taking into account this property and given that $c_{2,1}\left( \varphi
_{eff}\right) $ is assumed to be $k$-times piecewise continuously
differentiable, its Fourier series [30] can be approximated to leading
non-trivial order by a single harmonic function according to Table I (fifth
column). One sees that the universal scaling laws in Table I (fifth column)
yield $<V>=0$ when and only when both the shift symmetry and the
time-reversal symmetry of the system's dynamic equations (i.e., the force's
symmetry $f\left( -t\right) =f\left( t\right) $ in the present case) are
unbroken, while they yield a maximum value of $<V>$ when and only when
maximal \textit{effective} symmetry breaking occurs in the sense of the
measure (1), as predicted from the DSB mechanism. Also, that the harmonic
functions appearing in the universal scaling laws are independent of $har_{1}
$ is a consequence of the invariance of $<V>$ under time translation. As
expected, one finds that such universal scaling laws confirm and explain
previous results for a great diversity of systems [7,14,15,22,25] subjected
to a biharmonic force $f_{h1,h2}(t)$.

I now discuss how the aforementioned universal scaling laws change when the
violation of the time-reversal symmetry of the system's dynamic equations
cannot be absorbed in the temporal force. This is the case when dissipation
[5,6,8-13,16-21,24] is not negligible. It has been demonstrated above the
approximate conservation of the effective phase in the sense of the change
rules in Table I (fourth column), and hence that $\varphi _{eff}$ is the
proper argument of the function $c_{2,1}\left( \varphi _{1},\varphi
_{2}\right) $, when the violation of the time-reversal symmetry only occurs
in the temporal force. Therefore, that the violation of such a symmetry is
also due to the presence of dissipation means that $\varphi _{eff}$ can no
longer be an argument of the function $c_{2,1}\left( \varphi _{1},\varphi
_{2}\right) $ but one has $\varphi _{eff}+\varphi _{diss}$ instead, where $%
\varphi _{diss}$ is hereafter referred to as the \textit{dissipation phase}.
Note that the additive character of the dissipation phase is a consequence
of the DSB mechanism. Thus, the dissipation phase quantifies the degree of
breaking of the time-reversal symmetry generated by dissipation. Also, the
DSB mechanism implies the universal properties: $\varphi _{diss}\left( \beta
=0\right) =0$ and that $\varphi _{diss}\left( \beta \right) $ is a
monotonously increasing function of $\beta $, with $\beta $ being the
effective dissipation parameter. For the values of $\varphi _{eff}$ yielding 
$<V>=0$ in the absence of dissipation, i.e., those values for which the
temporal force does not break the time-reversal symmetry of a
non-dissipative system, one obtains that the maximum absolute value (i.e.,
1) of the harmonic functions appearing in the universal scaling laws is
reached at $\varphi _{diss}=\pm \pi /2$, and hence we have the additional
universal property $\max_{\beta }\varphi _{diss}\left( \beta \right) =\pi /2$
(cf. Table I, fifth column). However, the function $\varphi _{diss}\left(
\beta \right) $ generally depends upon additional parameters, such as the
period and diverse system-dependent parameters, i.e., it is not a universal
function. Of course, it is generally expected that the function $%
<V>/har\left( \varphi _{eff}+\varphi _{diss}\right) $ should exhibit
monotonously decreasing behaviour as a function of $\beta $, where $har$ is
the corresponding harmonic function in Table I (fifth column) in each case.
When dissipative forces dominate inertia (the so-called overdamped regime
[1,4]), the breaking of the time-reversal symmetry implies the breaking of
the force's symmetry $f\left( -t\right) =-f\left( t\right) $ and the
dissipative phase reaches its limiting values $\varphi _{diss}=\pm \pi /2$.
Since the optimal values of the relative amplitude $\epsilon _{2}/\epsilon
_{1}$ and the effective phase $\varphi _{eff}$ are just the same for the
three symmetries of the biharmonic force, this means that the universal
scaling laws corresponding to the overdamped regime are those given in Table
I (fifth column) but with $\sin $ instead of $\cos $, and vice versa, in
each case. One finds that these predictions are in perfect agreement with
published results for a great diversity of systems [4-6,8-13,16-21,24]
subjected to a biharmonic force $f_{h1,h2}(t)$. Since dissipative forces and
randomly fluctuating forces (noise) have the same microscopic origin, it is
expected the effectiveness of temporal forces at generating directed
transport induced by the ratchet effect to be robust against moderate
presence of noise.

In summary, universal scaling laws for the strength of directed transport
induced by symmetry breaking of temporal forces have been deduced from a
quantitative interpretation of Curie's principle. The present theory
explains in a general setting all previously published results for a great
diversity of systems [5-25], and provides a universal quantitative criterion
to optimize any application of the ratchet effect induced by symmetry
breaking of temporal forces.

\begin{acknowledgments}
The author warmly thanks N. R. Quintero for fruitful discussions. This study
was supported by the Spanish MCyT and the European Regional Development Fund
(FEDER) program through project FIS2004-02475.
\end{acknowledgments}

\bigskip

\textbf{TABLE I}

\begin{tabular}[t]{|c|c|c|c|c|}
\hline
$har_{1},har_{2}$ & $D_{s,\pm }$ & $t\rightarrow t+t_{0}$ & $t\rightarrow
-t+t_{0}$ & $\left\langle V\right\rangle $ \\ \hline
$\cos ,\cos $ & $\frac{\epsilon _{2}}{\epsilon _{1}}=1/2,\varphi
_{eff}=\left\{ \frac{\pi }{2},\frac{3\pi }{2}\right\} $ & $\widetilde{%
\varphi }_{eff}=\varphi _{eff}$ & $\widetilde{\varphi }_{eff}=-\varphi _{eff}
$ & $\sim \epsilon _{1}^{2}\epsilon _{2}\sin \varphi _{eff}$ \\ \hline
$\sin ,\sin $ & $\frac{\epsilon _{2}}{\epsilon _{1}}=1/2,\varphi
_{eff}=\left\{ 0,\pi \right\} $ & $\widetilde{\varphi }_{eff}=\varphi _{eff}$
& $\widetilde{\varphi }_{eff}=-\varphi _{eff}\pm \pi $ & $\sim \epsilon
_{1}^{2}\epsilon _{2}\cos \varphi _{eff}$ \\ \hline
$\sin ,\cos $ & $\frac{\epsilon _{2}}{\epsilon _{1}}=1/2,\varphi
_{eff}=\left\{ \frac{\pi }{2},\frac{3\pi }{2}\right\} $ & $\widetilde{%
\varphi }_{eff}=\varphi _{eff}$ & $\widetilde{\varphi }_{eff}=-\varphi _{eff}
$ & $\sim \epsilon _{1}^{2}\epsilon _{2}\sin \varphi _{eff}$ \\ \hline
$\cos ,\sin $ & $\frac{\epsilon _{2}}{\epsilon _{1}}=1/2,\varphi
_{eff}=\left\{ 0,\pi \right\} $ & $\widetilde{\varphi }_{eff}=\varphi _{eff}$
& $\widetilde{\varphi }_{eff}=-\varphi _{eff}\pm \pi $ & $\sim \epsilon
_{1}^{2}\epsilon _{2}\cos \varphi _{eff}$ \\ \hline
\end{tabular}

\bigskip

Table I. Optimal values of the relative amplitude $\epsilon _{2}/\epsilon
_{1}$ and the effective phase $\varphi _{eff}\equiv \varphi _{2}-2\varphi
_{1}$ obtained by computing the measure (1) of DSB for the three symmetries
of the biharmonic force (second column), change rules of the effective phase
under time transformations (third and fourth columns), and universal scaling
laws in leading order for averaged velocities and currents. Note the
coherence of the results in the second and fifth columns, which were
obtained using independent methods (recall that, without loss of generality, 
$\epsilon _{1}+\epsilon _{2}=1$ so that $\epsilon _{1}^{2}\epsilon
_{2}=\left( 1-\epsilon _{2}\right) ^{2}\epsilon _{2}$, which is a function
having a single maximum at $\epsilon _{2}=1/3,$ and hence $\epsilon
_{1}=2/3,\epsilon _{2}/\epsilon _{1}=1/2.$)

\bigskip

\bigskip

{\Large Figure Captions}

\bigskip

Figure 1. (a) Elliptic force $f_{ellip}(t)=\epsilon f(t;T,m,\theta )\equiv
\epsilon \limfunc{sn}\left( \Omega t+\Theta ;m\right) \limfunc{cn}\left(
\Omega t+\Theta ;m\right) $ vs $t/T$ and three shape parameter values, $m=0$
(light blue line), $m=0.983417$ (dark blue line, optimal universal
waveform), $m=1-10^{-6}$(light blue line), showing an increasing
symmetry-breaking sequence as the pulse narrows, i.e., as $m\rightarrow 1$.
(b) Optimal universal biharmonic force generating directed transport in one
direction ($f_{bihar}^{+}\left( t\right) /\epsilon =\sin (2\pi t/T)+\frac{1}{%
2}\sin \left( 4\pi t/T\right) $, dark blue line) and the opposite direction (%
$f_{bihar}^{-}\left( t\right) /\epsilon =\sin (2\pi t/T)-(1/2)\sin \left(
4\pi t/T\right) \equiv -f_{bihar}^{+}\left( t+T/2\right) /\epsilon $, red
line). (c) Measure of the DSB (Eq. (1)) for the elliptic force in (a), $%
D_{s}(f_{ellip})=E(m)K^{-1}(m)\left( 1-m\right) ^{-1/2}$ vs $m$. One sees a
sharp increase as $m\rightarrow 1$. (d) Measure of the DSB (Eq.(1)) for the
biharmonic force $f_{bihar}\left( t\right) /\epsilon =\cos \left( 2\pi
t/T\right) +a\cos \left( 4\pi t/T+\varphi _{eff}\right) $, $D_{s}(f_{bihar})$
vs $a$ for $\varphi _{eff}=\pi /2$. One sees a sharp increase as $%
a\rightarrow 1/2$, which is similar to that found in (c) for the elliptic
function.


\bigskip 

\begin{thebibliography}{99}
\bibitem{1} R. P. Feynman, R. B. Leighton, and M. Sands, \textit{The Feynman
Lectures on Physics} (Addison-Wesley, Reading, Massachusetts, 1966),Vol. 1,
Ch. 46.

\bibitem{2} P. H\"{a}nggi and R. Bartussek, in \textit{Lecture Notes in
Physics}, Vol. 476, p. 294, edited by J. Parisi \textit{et al}. (Springer,
Berlin, 1996).

\bibitem{3} Special issue on \textit{Ratchets and Brownian Motors: Basics,
Experiments and Applications}, edited by H. Linke [Appl. Phys. A \textbf{75}
(2002)].

\bibitem{4} P. Reimann, Phys. Rep.\textit{\ }\textbf{361}, 57 (2002).

\bibitem{5} K. Seeger and W. Maurer, Solid State Commun. \textbf{27}, 603
(1978).

\bibitem{6} A. Ajdari, D. Mukamel, L. Peliti, and J. Prost, J. Phys. I
France \textbf{4}, 1551 (1994).

\bibitem{7} E. Dupont, P. B. Corkum, H. C. Liu, M. Buchanan, and Z. R.
Wasilewski, Phys. Rev. Lett. \textbf{74}, 3596 (1995).

\bibitem{8} K. N. Alekseev, M. V. Erementchouk, and F. V. Kusmartsev,
Europhys. Lett.\textit{\ }\textbf{47}, 595 (1999).

\bibitem{9} S. Flach, O. Yevtushenko, and Y. Zolotaryuk, Phys. Rev. Lett.%
\textit{\ }\textbf{84}, 2358 (2000).

\bibitem{10} O. Yevtushenko, S. Flach, Y. Zolotaryuk, and A. A. Ovchinnikov,
Europhys. Lett.\textit{\ }\textbf{54}, 141 (2001).

\bibitem{11} M. Salerno and Y. Zolotaryuk, Phys. Rev. E \textbf{65}, 56603
(2002).

\bibitem{12} S. Flach, Y. Zolotaryuk, A. E. Miroshnichenko, and M. V.
Fistul, Phys. Rev. Lett. \textbf{88}, 184101 (2002).

\bibitem{13} A. Engel, H. W. M\"{u}ller, P. Reimann, and A. Jung, Phys. Rev.
Lett.\textit{\ }\textbf{91}, 60602 (2003).

\bibitem{14} M. Schiavoni, L. S\'{a}nchez-Palencia, F. Renzoni, and G.
Grynberg, Phys. Rev. Lett.\textit{\ }\textbf{90}, 94101 (2003).

\bibitem{15} M. V. Fistul, A. E. Miroshnichenko, and S. Flach, Phys. Rev. B%
\textit{\ }\textbf{68}, 153107 (2003).

\bibitem{16} J. Lehmann, S. Kohler, P. H\"{a}nggi, and A. Nitzan, J. Chem.
Phys.\textit{\ }\textbf{118}, 3283 (2003).

\bibitem{17} L. Morales-Molina, N. R. Quintero, F. G. Mertens, and A. S\'{a}%
nchez, Phys. Rev. Lett.\textit{\ }\textbf{91}, 234102 (2003).

\bibitem{18} A. V. Ustinov, C. Coqui, A. Kemp, Y. Zolotaryuk, and M.
Salerno, Phys. Rev. Lett.\textit{\ }\textbf{93}, 87001 (2004).

\bibitem{19} R. Gommers, S. Bergamini, and F. Renzoni, Phys. Rev. Lett.%
\textit{\ }\textbf{95}, 73003 (2005).

\bibitem{20} P. H\"{a}nggi, F. Marchesoni, and F. Nori, Ann. Phys. (Leipzig)%
\textit{\ }\textbf{14}, 51 (2005).

\bibitem{21} X. Xie, J. Dai, and X.-C. Zhang, Phys. Rev. Lett.\textit{\ }%
\textbf{96}, 75005 (2006).

\bibitem{22} I. Franco and P. Brumer, Phys. Rev. Lett.\textit{\ }\textbf{97}%
, 40402 (2006).

\bibitem{23} D. Cole \textit{et al}. Nature Mat. \textbf{5}, 305 (2006).

\bibitem{24} Y. Zolotaryuk and M. Salerno, Phys. Rev. E\textit{\ }\textbf{73}%
, 66621 (2006).

\bibitem{25} S. Denisov, L. Morales-Molina, and S. Flach,
arXiv:cond-mat/0607558 (2006).

\bibitem{26} P. Curie, J. de Phys. (Paris) \textbf{3}, 393 (1894).

\bibitem{27} D. Zwillinger, \textit{Standard Mathematical Tables and Formulae%
} (Chapman \& Hall/CRC, London, 2003).

\bibitem{28} R. Chac\'{o}n and N. R. Quintero, BioSystems (to be published).

\bibitem{29} M. Newman, Nature \textbf{405}, 412 (2000); J. M. Carlson and
J. Doyle, Phys. Rev. Lett.\textit{\ }\textbf{84}, 2529 (2000).

\bibitem{30} Y. Katznelson, \textit{An Introduction to Harmonic Analysis}
(Dover, New York, 1976).
\end{thebibliography}
\end{document}